\documentclass[iop, apj, numberedappendix, twocolappendix]{emulateapj}
\usepackage{times}
\usepackage{ amssymb }

\usepackage{hyperref}
\usepackage{xcolor}
\usepackage{amsmath}
\definecolor{dark-red}{rgb}{0.4,0.15,0.15}
\definecolor{dark-blue}{rgb}{0.15,0.15,0.4}
\definecolor{medium-blue}{rgb}{0,0,0.5}
\hypersetup{
    colorlinks,
    linkcolor={blue},
    citecolor={blue},
    urlcolor={blue}
}
\usepackage[varg]{txfonts}

\usepackage{color}

\newcommand{\para}{\parallel}

\newcommand{\paren}[1]{ \left( #1 \right) }
\newcommand{\kb}{k_{\rm B}}

\newcommand{\dmu}{d\ln \mu/d\ln P}
\newcommand{\dT}{d\ln T/d\ln P}
\renewcommand{\b}{\hat{\bb{b}}}

\newcommand{\be}{\begin{eqnarray}}
\newcommand{\en}{\end{eqnarray}}
\newcommand{\pa}{\partial}
\newcommand{\f}{\frac}

\newcommand{\oma}{\omega_{\rm A}}
\newcommand{\dvx}{\delta v_{x}}
\newcommand{\dvz}{\delta v_{z}}
\newcommand{\dbx}{\delta B_{x}}
\newcommand{\dbz}{\delta B_{z}}
\newcommand{\dt}{\delta T/T}
\newcommand{\drho}{\delta \rho/\rho}
\newcommand{\dm}{\delta \mu/\mu}
\newcommand{\mH}{m_{\rm H}}
\newcommand{\vtho}{v_{\mathrm{th}, 0}}

\newcommand{\omdy}{\omega_{\rm dyn}}

\newcommand{\D}[2]{\frac{\partial #1}{\partial #2}}
\newcommand\bb[1]{\mbox{\boldmath{$#1$}}}
\newcommand\bcdot{\bb{\cdot}}
\newcommand\del{\nabla}
\newcommand\btimes{\times}

\begin{document}

\title{Local Simulations of Instabilities Driven by Composition Gradients in the ICM}

\author{Thomas Berlok and Martin~ E. Pessah}
\affil{Niels Bohr International Academy, Niels Bohr Institute, Blegdamsvej
17, DK-2100 Copenhagen \O, Denmark;
\url{berlok@nbi.dk, mpessah@nbi.dk}}

\shorttitle{Local Simulations of Instabilities Driven by Composition Gradients in the ICM}

\shortauthors{Berlok \& Pessah}

\begin{abstract}
The distribution of Helium in the intracluster medium (ICM) permeating galaxy
clusters is not well constrained due to the very high plasma temperature.
Therefore, the plasma is often assumed to be homogeneous. A non-uniform Helium
distribution can however lead to biases when measuring key cluster parameters.
This has motivated one-dimensional models that evolve the ICM composition
assuming that the effects of magnetic fields can be parameterized or ignored.
Such models for non-isothermal clusters show that Helium can sediment in the
cluster core leading to a peak in concentration offset from the cluster
center. The resulting profiles have recently been shown to be linearly
unstable when the weakly-collisional character of the magnetized plasma is
considered. In this paper, we present a modified version of the MHD code
Athena, which makes it possible to evolve a weakly-collisional plasma subject
to a gravitational field and stratified in both temperature and composition.
We thoroughly test our implementation and confirm excellent agreement against
several analytical results.  In order to isolate the effects of composition,
in this initial study we focus our attention on isothermal plasmas. We show
that plasma instabilities, feeding off gradients in composition, can induce
turbulent mixing and saturate by re-arranging magnetic field lines and
alleviating the composition gradient. Composition profiles that increase with
radius lead to instabilities that saturate by driving the average magnetic
field inclination to roughly $45^{\circ}$. We speculate that this effect may
alleviate the core insulation observed in homogeneous settings, with potential
consequences for the associated cooling flow problem.
\end{abstract}


\section{Introduction}

Atmospheres comprised of a plasma that is weakly collisional and weakly
magnetized have stability properties that differ qualitatively from
collisional atmospheres. Instabilities such as the magneto-thermal instability
(MTI, \citealt{Bal00,Bal01}) and the Heat-Flux-driven Buoyancy Instability
(HBI, \citealt{Qua08}) can arise when there is a gradient in the temperature
either parallel or anti-parallel to the gravitational field. These
instabilities, that feed off a gradient in temperature, have been extensively
studied
\citep{Bal00,Bal01,Qua08,Kun11,Lat12,Par05,Par07,Par08,Par08_MTI,Par09,Bog09,
2010ApJ...712L.194P,2010ApJ...713.1332R,2011MNRAS.413.1295M,
2012MNRAS.419.3319M,Kun12,Par12,2012MNRAS.419L..29P} and they are believed to
be important for the understanding of the dynamical evolution of the
intracluster medium (ICM) of galaxy clusters.

These studies assumed that the composition of the plasma is uniform,  an
assumption which might not be appropriate if heavier elements are able to
sediment towards the core of the cluster \citep{Fab77}. In parallel and
complementary studies, the long-term evolution of the radial distribution of
elements has been studied using one-dimensional models
\citep{Fab77,Gil84,Chu03,Chu04,Pen09,Sht10}. The ensuing non-uniform
composition has been argued to introduce biases in cluster properties as
inferred from observations \citep{Mar07,Pen09}.

While the studies of the MTI and HBI assumed a uniform plasma the
sedimentation models have yet to include magnetic fields. In an attempt to
bridge the gap between the different approaches, and with the goal of
understanding the long-term evolution of the composition of the ICM,
\cite{Pes13} studied the stability properties of weakly collisional
atmospheres with gradients in both temperature and composition. They found
that gradients in composition, either parallel or anti-parallel to the
gravitational field, can trigger instabilities. In a subsequent study,
\cite{Ber15a} carried out a comprehensive study using linear mode analysis and
showed that these instabilities are expected to render the composition
profiles obtained with current sedimentation models unstable, as it was
illustrated using the model of \cite{Pen09}.

In this paper, we present the first nonlinear, two-dimensional (2D),
numerical simulations of the instabilities that feed off a gradient in
composition using a modified version of the MHD code Athena
\citep{stone_athena:_2008}. The instabilities considered are \emph{i)} the
Magneto-Thermo-Compositional Instability (MTCI) which is maximally unstable
when the magnetic field is perpendicular to gravity \emph{ii)} the Heat- and
Particle-flux-driven Buoyancy Instability (HPBI) which is maximally unstable
when the magnetic field is parallel to gravity and \emph{iii)} the diffusion
modes which are maximally unstable when the magnetic field is parallel to
gravity. These instabilities arise due to the weakly collisional
nature of the ICM, which fundamentally changes the transport properties of a
plasma. In this regime, where the gyro-radii of the particles are much smaller
than the mean free path for particle collisions, the transport of heat,
momentum and particles will be primarily along the magnetic field lines.

The MTCI and HPBI will be present in isothermal atmospheres in which the
composition increases with height while diffusion modes can be present
regardless of the direction of the gradient in composition \citep{Pes13}. The
linear dispersion relation presented in \cite{Ber15a} is used to compare with
the linear evolution of the simulations. We find good agreement thereby
confirming both the linear theory and our numerical method. For the nonlinear
evolution of the instabilities we find that the magnetic field inclination
goes to roughly $45^{\circ}$ independently of whether the magnetic field is
initially horizontal (MTCI) or vertical (HPBI). This is contrary to the
instabilities driven by temperature gradients where the average magnetic field
becomes almost vertical (horizontal) for an initially horizontal (vertical)
magnetic field \citep{Par05,Par08}. The simple explanation is that the MTCI
and HPBI, both of which grow when the the composition increases with height,
can operate simultaneously. They are therefore driving the average angle in
opposite directions, compromising at roughly $45^{\circ}$. The MTI and HBI,
being dependent on temperature gradients in opposite directions, cannot grow
at the same time and so they grow unabated by their counterpart. We also find
that both types of instabilities cause turbulent mixing of the Helium
concentration. We conclude that, in the idealized numerical settings that we
employ, instabilities driven by the free energy supplied by a gradient in
composition  saturate by alleviating the gradient and thereby removing the
source of free energy.

The rest of the paper is organized as follows: We start out by introducing the
equations of kinetic MHD in Section \ref{sec:Kinetic MHD for a binary mixture}
and how they can be solved numerically in Section \ref{sec:Numerical method
and initial conditions}. In section \ref{sec:Simulations of the Linear Regime}
we demonstrate that the simulations agree with the linear theory for
isothermal atmospheres and we illustrate how the growth rates depend on some
of the key parameters of the problem. We also use atmospheres with gradients
in both temperature and composition, motivated by the model of \cite{Pen09}
and discussed in \cite{Ber15a}, to show that the theory and simulations also
agree with both gradients present. In Section \ref{sec:Simulations of the
Nonlinear Regime}, we consider the nonlinear evolution of the MTCI and HPBI in
isothermal atmospheres in order to determine how they saturate. We summarize
and outline future work in Section \ref{sec:Summary and discussion}.

\section{Kinetic MHD for a binary mixture}
\label{sec:Kinetic MHD for a binary mixture}

We consider a fully ionized, weakly magnetized, and weakly collisional plasma
consisting of a mixture of Hydrogen and Helium. We model such a plasma using
the set of equations introduced in \citet{Pes13}\footnote{For further details
on the kinetic MHD approximation and its limitations see the
relevant discussions in \citet{Kun12,schekochihin_plasma_2005,Pes13} and
references therein.}
\begin{eqnarray}
\f{\pa \rho}{\pa t}+\del\bcdot(\rho \bb{v}) &=& 0 \,,
\label{eq:rho}\\
\D{\paren{\rho \bb{v}}}{t}
+\del \bcdot \paren{\rho \bb{vv} + P_{\mathrm{T}} \mathsf{I} - \frac{B^2}{4\pi} \b\b}
&=&
-\del \bcdot \Pi
+ \rho \bb{g}  , \, \\
\D{E}{t} + \del \bcdot \left[ \paren{E+P_{\mathrm{T}}}\bb{v} -
\frac{\bb{B}\paren{\bb{B\cdot v}}}{4\pi}\right]
&=& -\del \bcdot \bb{Q}_{\rm s}
- \del \bcdot
\paren{\Pi \bcdot \bb{v}} +\rho \bb{g\cdot v} \ ,
\label{eq:E}  \nonumber \\ \\
\f{\pa \bb{B}}{\pa t}&=&\del\btimes(\bb{v}\btimes\bb{B}) \,,
\label{eq:b}\\
\D{\paren{c\rho}}{t}+\del\bcdot(c \rho \bb{v}) &=&-\del\bcdot\bb{Q}_{\rm c} \,.
\label{eq:c}
\end{eqnarray}
In these equations $\rho$ is the mass density, $\bb{v}$ is the
fluid velocity, $\bb{B}$ is the magnetic field with direction $\hat{\bb{b}} = (b_x, 0, b_z)$, $\bb{g} = (0, 0, -g)$ is the gravitational acceleration and
$\mathsf{I}$ is the identity matrix. The total pressure is
$P_{\mathrm{T}} = P+ {B^2}/{8  \pi}$ where $P$ is the thermal pressure and the total energy density,
$E$, is
\be
E = \f{1}{2}\rho v^2 + \f{B^2}{8\pi} + \f{P}{\gamma - 1} \ ,
\en
where $\gamma =5/3$ is the adiabatic index.

The composition of the plasma, $c$,
is defined to be the ratio of the Helium density
to the total gas density
\be
c \equiv \f{\rho_{\rm He}}{\rho_{\rm H} + \rho_{\rm He}} =
\f{\rho_{\rm He}}{\rho} \ ,
\en
and the associated mean molecular weight, $\mu$, is given by
\be
\mu=\frac{4}{8-5c} \ , \label{eq:relation_from_c_to_mu}
\en
for a completely ionized plasma consisting of Helium
and Hydrogen.
The mean molecular weight can modify the dynamics of the plasma through the equation of state
\be
P= \frac{\rho k_{\rm B} T}{\mu m_{\rm H}} \, , \label{eq:eos}
\en
where $\kb$ is Boltzmann's constant, $T$ is the temperature and $m_{\rm H}$ is
the proton mass.

We consider the plasma to be influenced by three different non-ideal
effects:
Braginskii viscosity, which arises due to differences in pressure parallel
($p_\para$) and perpendicular ($p_\perp$) to the magnetic field, described by the viscosity tensor \citep{Bra}
\be
\Pi = - 3\rho \nu_\para \paren{\b\b -\f{1}{3} \mathbf{I} } \paren{ \b\b -
\frac{1}{3}\mathbf{I} }  \bb{:} \del \bb{v}  \ ,
\en
anisotropic heat conduction  described by the heat flux
\citep{1962pfig.book.....S,Bra}
\be
\bb{Q}_{s}=-\chi_\para\hat{\bb{b}}\hat{\bb{b}}\cdot\nabla T,
\label{eq:heat-flux}
\en
and anisotropic diffusion of composition described by the composition flux
\citep{1990ApJ...360..267B}
\be
\bb{Q}_{c}=-D\hat{\bb{b}}\hat{\bb{b}}\cdot\nabla c \,.
\label{eq:helium-flux}
\en
The transport coefficients for Braginskii viscosity ($\nu_{\para}$), heat
conductivity ($\chi_\para$) and diffusion of composition ($D$) depend on the
temperature, density and composition of the plasma. The dependences are given
by Equations (64)-(66) in \cite{Ber15a}. Finally, we define the thermal
velocity, $v_{\rm th} = \sqrt{P/\rho}$ and the plasma-$\beta$ given by $ \beta
=  8\pi P/B^2 = 2{v_{\rm th}^{2}}/{v_{\rm A}^{2}}, $ where $v_{\rm A}=
{B}/\sqrt{4 \pi \rho}$ is the Alfv{\'e}n velocity\footnote{Note that this
definition of $\beta$ differs from the one in \cite{Ber15a} by a factor of
2.}.

\section{Numerical method and initial conditions}
\label{sec:Numerical method and initial conditions}

The equations of kinetic MHD, Equation (\ref{eq:rho})-(\ref{eq:c}), are solved
using a modified version of the conservative MHD code Athena
\citep{stone_athena:_2008}. The algorithms  used in Athena are described in
\cite{gardiner_unsplit_2005, stone_simple_2009} and a description of the
implementation of anisotropic thermal conduction and Braginskii viscosity can
be found in \cite{Par05} and \cite{Par12}, respectively.

In order to carry out the numerical simulations of interest,
we have modified Athena to include a spatially varying mean molecular weight,
$\mu$. This is done by using the inbuilt method for adding a passive scalar,
defined by a spatially varying concentration, $c$, and then making it active
by using the value of $c$ when calculating the temperature used in the heat
conduction module. Furthermore, we implemented a module which takes account of
diffusion of Helium by using operator splitting. This module has been built
by following the same approach employed in the heat conduction module
that is already present in the current publicly available version of Athena
\citep{Par05,Sha07}. Our implementation allows for non-constant values of the
parameters $\nu_\para$, $\chi_\para$,  and $D$ through user-defined functions
This feature is, however, not used in this work, as we employ
a local approximation and thus treat these parameters as constants. The diffusion
terms are solved explicitly which can make the
time-step constraint on viscosity, thermal conduction, and diffusion of Helium
very restrictive. In order to circumvent this we use subcycling, which we
limit to a maximum of ten steps per MHD step \citep{Kun12}.

\subsection{Plane-parallel atmosphere with gradients in temperature and
composition}
\label{sec:Plane-parallel atmosphere with gradients in temperature and
composition}

In this section, we introduce the two different atmospheres used as initial
conditions in the simulations. The atmospheres considered are plane-parallel,
i.e., all quantities are constant along a horizontal slice, perpendicular to gravity.
The atmosphere is assumed to be composed of an ideal gas, characterized
by the equation of state given by Equation (\ref{eq:eos}), and is assumed to
be in hydrostatic equilibrium, i.e.,\footnote{We consider high-$\beta$ plasmas
and do not include the magnetic pressure in the derivations of the equilibria.}
\be
\D{P}{z}=-g\rho \,. \label{eq:hydrostatic_equilibrium}
\en

\subsubsection{Isothermal Atmosphere with a Composition Gradient
in the Absence of Particle Diffusion}
\label{sec:simple_atmo}

The simplest atmosphere we use is inspired by the original numerical work on
the MTI \citep{Par05}. We consider an isothermal atmosphere with $T = T_{0}$ and
\be
P & = & P_{0}\left(1-\frac{z}{3H_{0}}\right)^{3} \ ,\\
\mu & = & \mu_{0}\left(1-\frac{z}{3H_{0}}\right)^{-1} \ .
\en
where $P_{0}$, $T_{0}$, and $\mu_{0}$ are the values of the pressure,
temperature, and mean molecular weight at $z = 0$, and $H_0$ is the
scale height
\be
H_0 = \frac{\kb T_0}{\mu_0 \mH g} \ .
\en
The density can be determined using Equation (\ref{eq:eos}).

This isothermal atmosphere is used for simulations of the linear regime of the
MTCI in Section \ref{sec:The Magneto-Thermo-Compositional Instability} and the
linear regime of the HPBI in Section
\ref{sec:The Heat and Particle-flux-driven Buoyancy Instability}. It is also
used for simulations of the nonlinear
regime of the MTCI and HPBI in Section \ref{sec:Simulations of the Nonlinear
Regime}. The magnetic field can have any orientation as long as $D = 0$. The
structure of this atmosphere is however not in equilibrium
if $D\neq 0$ and $b_z \neq 0$. In that case, we will have to use a more
sophisticated atmosphere which we introduce next.

\subsubsection{Atmosphere with Thermal and Composition Gradients}
\label{sec:advanced_atmo}

Steady state requires that the divergence of the heat and particle fluxes
vanish, i.e.,
\be
\del \bcdot {\bb{Q}_{\rm s}} &=& 0 \ , \\
\del \bcdot {\bb{Q}_{\rm c}} &=& 0\ ,
\en
Both conditions are trivially satisfied if $b_z = 0$ and $D = 0$. If, however,
$b_z \ne 0$ and $D \neq 0$, these requirements can still be met by
simple atmospheric models if $\chi_\para$ and $D$ do not depend on $z$.
Such an assumption is reasonable
for the local simulations that we will consider,
where the height of the box, $L_z$, satisfies the
criterion $L_z \ll H_0$. When there is both a gradient in temperature $T$ and
mean molecular weight $\mu$, the requirements that $\del \bcdot {\bb{Q}_{\rm s}} = 0$
and $\del \bcdot {\bb{Q}}_{\rm {c}} = 0$ can be integrated to yield
\be
T(z) &=& T_{0}+ s_{\rm T} z \ , \\
c(z) &=& c_{0}+ s_{\rm c} z \ ,
\en
where $s_{\rm T} = (T_{\rm Z}-T_0)/L_{\rm Z}$ and $s_{\rm c} = (c_{\rm
Z}-c_0)/L_{z}$ are the constant slopes
in temperature and composition. Here, $T_0$ ($T_{\rm Z}$) is the temperature at the
bottom (top) of the box and $c_0$ ($c_{\rm Z}$) is the Helium mass concentration at the
bottom (top) of the box.

The pressure is found by solving Equation (\ref{eq:hydrostatic_equilibrium}),
leading to
\be
P(z) = P_0 \left( \f{T(z) \mu(z)}{T_0 \mu_0}\right)^{\alpha} \ ,
\en
where $\mu(z)$ is related to $c(z)$ by Equation
(\ref{eq:relation_from_c_to_mu}) and the constant $\alpha$ is given by
\be
\alpha = - \f{T_0}{H_0} \f{4}{4s_{\rm T} + 5 \mu_0 T_0 s_{\rm c}} \ .
\en
This solution for the pressure profile of the atmosphere is replaced with a
simple exponential atmosphere, $P(z) = P_0 \exp(-z/H_0)$, with scale height
$H_0$ if $s_{\rm T} = s_{\rm c} = 0$.

We use this model atmosphere to perform simulations of modes driven by diffusion in
Section \ref{sec:Modes Driven by Diffusion}. These modes are unstable when
there is a vertical gradient in composition, a non-zero vertical component of
the magnetic field, $b_z \neq 0$, and anisotropic diffusion of Helium, $D \neq
0$. We also use this atmosphere in Section \ref{sec:Gradients in temperature
and composition} for simulations of the linear regime of the MTCI and the HBPI
with gradients in both temperature and composition.

\subsection{Boundary conditions}
\label{sec:Boundary conditions}

Periodic boundary conditions are used in the horizontal direction in all
simulations. In the vertical direction we have implemented two different sets
of boundary conditions \emph{i)} the conventional reflective boundary
conditions and \emph{ii)} a set of boundary conditions that we will call
quasi-periodic boundary conditions. Both sets of boundary conditions are
explained in detail in Appendix \ref{sec:Boundary conditions_appendix}. Here,
we give a brief account of the motivation for using these two sets of boundary
conditions and their key differences.

\emph{The quasi-periodic boundary conditions}  are periodic in the relative
changes in the physical quantities. We have found that these boundary
conditions are a necessity in order for the simulations to reproduce the
growth rates predicted by the local linear mode analysis. We believe that this is
due to the assumption of periodicity in the perturbed quantities that is made
when the dispersion relation is derived. This problem has also been
encountered in previous studies of the MTI \citep{Ras08}. These boundary conditions
are used in all simulations presented in Section \ref{sec:Simulations of the
Linear Regime}.

\emph{The reflective boundary conditions} maintain hydrostatic equilibrium
by extrapolating pressure and density into the ghost zones at the top and
bottom of the computational domain. The values of temperature and composition
are held fixed at their initial values in the ghost zones. The velocity
$z$-component is reflected symmetrically around the boundaries. If the
magnetic field is initially vertical (horizontal) it is forced to remain
vertical (horizontal) at the boundaries. These boundary conditions are used in
the simulations presented in Section \ref{sec:Simulations of the Nonlinear
Regime}.

\section{Simulations of the Linear Regime}
\label{sec:Simulations of the Linear Regime}

The equations are made dimensionless by scaling the density with $\rho_0$,
distances with $H_0$, and velocities with the thermal velocity $v_{\rm th, \,
0}$. The magnetic field strength $B_0$ is found from the dimensionless
parameter $\beta_0$. Here, the subscript $"0"$ denotes
the value at the bottom of the computational domain, $z = 0$. With this
convention, the unit of time is $H_0/\vtho$,
temperature is scaled with $T_0$, $\mu$ is scaled with $\mu_0$,
pressure, as well as energy density, is scaled with $P_0 = \rho_0 \vtho^2$,
and the value of $g$ is unity. As a consequence, the coefficient for
anisotropic heat conduction, $\chi_\para$, is scaled with $\rho_0 \vtho^
3 H_0/T_0$
and the coefficients for Braginskii viscosity$, \nu_\para$, and anisotropic
diffusion of composition, $D$, are both scaled with $\vtho H_0$.

\begin{figure}[t]
\centering
\includegraphics{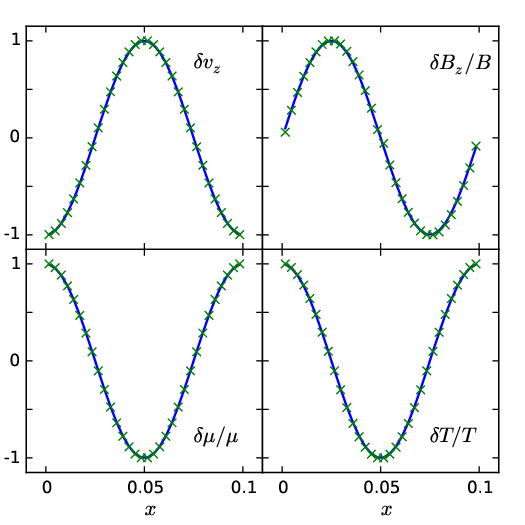}
\caption{Four of the components of the perturbation at $z = L_z/2$ for a mode
with $k_x = 2 \pi /L_x$ at time $t = 5$ in a simulation with resolution
$32 \times 32$. The simulation (green crosses) matches the theory
(blue lines). The magnetic field is $\pi/2$ out of phase with the velocity
perturbation, as expected for a purely growing mode.}
\label{fig:eigenmodes}
\end{figure}

We begin by comparing the simulations with the linear theory. In order to do
so, we use the quasi-periodic boundary conditions described in the previous
section and in Appendix \ref{sec:Boundary conditions_appendix}. A Cartesian
box of size $[0, L_x] \times [0, L_z]$ with $L_x = L_z = 0.1 $ and a
resolution of $64\times64$ is used in all simulations unless otherwise noted.
An overview of the simulations of the linear regime can be found
in Table~\ref{tbl-1}.
\begin{deluxetable*}{lcrcrcccc}[t]
\tabletypesize{\scriptsize}
\tablecaption{Simulations of the linear regime using
quasi-periodic boundary conditions. Each row represents a series of simulations
where the ellipses denote that the associated parameter is being varied.
\label {tbl-1}}
\tablewidth{0pt}
\tablehead{
\colhead{Simulation} & \colhead{($n_x,n_z$)} &\colhead{$\theta$}&\colhead
{$\beta_0$} &
\colhead{$\chi_\para$} & \colhead{$\nu_\para$} & \colhead{$D$} &\colhead{Resolution} & \colhead
{Figure}
}
\startdata
MTCI\_chi  & $(1,0)$ & $ 0^{\circ}$  & $2\cdot10^8$  &    \ldots & 0 & 0  & $64\times64$& \hyperref[fig:local_mtci]{\ref{fig:local_mtci}.a} \\
MTCI\_B    & $(1,0)$ & $ 0^{\circ}$  & \ldots  & $3\cdot10^{-4}$ & 0 & 0  &
$64\times64$&\hyperref[fig:local_mtci]{\ref{fig:local_mtci}.b}\\
HPBI\_nu   & $(1,1)$ & $90^{\circ}$  & $2\cdot10^8$  & $10^{-4}$ & \ldots & 0  &
$64\times64$&\hyperref[fig:local_hbci]{\ref{fig:local_hbci}.a}\\
HPBI\_n    & (\ldots, \ldots)  & $90^{\circ}$  & $2\cdot10^6$  & $10^{-4}$     & 0 & 0 &
$256\times256$&\hyperref[fig:local_hbci]{\ref{fig:local_hbci}.b} \\
D-mode\_D  & $(1,1)$ & $90^{\circ}$  & $2\cdot10^8$  & $10^{-3}$     & 0 & \ldots &
$256\times256$&\hyperref[fig:difmode1]{\ref{fig:difmode1}.a} \\
D-mode\_nu & $(1,1)$ & $90^{\circ}$  & $2\cdot10^8$  & $10^{-3}$ & \ldots & $10^
{-3}$  & $64\times64$&\hyperref[fig:difmode1]{\ref{fig:difmode1}.b} \\
MTCI\_ICM\tablenotemark{a} & (\ldots, 0) & $ 0^{\circ}$ & $2\cdot10^6$  &
$1.4\cdot10^{-2} $ & $4.0\cdot10^{-4}$ & $0$  & $256\times32$&
\hyperref[fig:icm_linear]{\ref{fig:icm_linear}.a} \\
HPBI\_ICM\tablenotemark{a} & (\ldots, \ldots) & $90^{\circ}$ & $2\cdot10^6$  &
$2.7\cdot10^ {-4}$ & $4.5\cdot10^{-6}$ & $0$ & $256\times256$&
\hyperref[fig:icm_linear]{\ref{fig:icm_linear}.b} \\
\enddata \tablenotetext{a}{Using gradients in both temperature and composition.}
\end{deluxetable*}

The instabilities are excited by seeding a given mode, with components
($\dvx$, $\dvz$, $\dbx$, $\dbz$, $\drho$, $\dt$, $\dm$), as derived by solving
the eigenvalue system associated with the dispersion relation
introduced in
\citet{Pes13,Ber15a}. We set the overall mode amplitude by enforcing $\drho =
10^{-4}$, so that the velocity perturbation is subsonic \citep{Par05}. The
amplitudes of the other components are fixed by the solution to the linear
eigenvalue problem, which predicts that unstable modes grow exponentially as
$\exp (\sigma t)$ while the ratio of their components remains constant in
time.

We begin by considering $D = 0$ and the hydrostatic atmosphere
given in Section \ref{sec:simple_atmo}, that has $\dmu = -1/3$ and $\dT = 0$.
This atmosphere is unstable regardless
of whether the magnetic field is oriented horizontally (MTCI) or vertically
(HPBI), as described in \cite{Ber15a}.

\begin{figure}[t]
\centering
\includegraphics{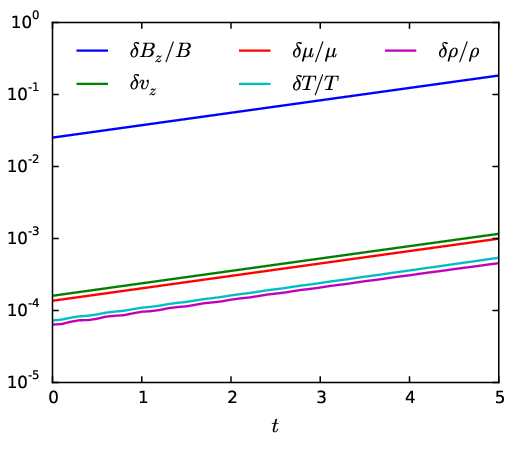}
\caption{Evolution of box-averaged quantities. The perturbed quantities grow
exponentially with a growth rate $\sigma = 0.40$.}
\label{fig:perturbed_evolution}
\end{figure}

\subsection{The Magneto-Thermo-Compositional Instability}
\label{sec:The Magneto-Thermo-Compositional Instability}

When the magnetic field is perpendicular to gravity the general dispersion
relation, Equation (13) in \cite{Ber15a}, reduces to
\be
\sigma^{2}\approx-g\frac{d\ln\left(T/\mu\right)}{dz}\frac{k_{x}^{2}+
k_{y}^{2}}{k^{2}}
\label{eq:approx_sig_MTCI} \ ,
\en
in the limit of fast heat conduction and weak magnetic field. When $\mu$
increases with height and the atmosphere is isothermal we have $\sigma > 0$.
This is the instability known as the MTCI \citep{Pes13}. In order to excite a
single MTCI mode, we use a perturbation of the form\footnote{We note that
$k_y = 0$ in all the simulations presented in this paper.} $k_z = 0 $
and $k_x = 2 \pi/L_{x}$.
We are interested in a direct visual comparison of the spatial dependence of
the perturbations in the simulations and the one expected from the linear
theory. In order to illustrate this, we consider a setting with
$\chi_{\para} = 3\cdot 10^{-4}$ and $\beta_0 = 2\cdot 10^{8}$. In Figure
\ref{fig:eigenmodes}, we show the values of  the perturbations (green crosses)
$\dvz$, $\dbz/B$, $\delta \mu/\mu$ and $\delta T/T$ as a function of the
$x$-coordinate. The
data slices are drawn at a fixed height, $z = L_{z}/2$ at the time $t = 5$ in
dimensionless units. The numerical results show good agreement with the
analytical results shown with blue solid lines.

In order to calculate the growth rate of the mode, we perform an exponential
fit to the time evolution of the box average of the absolute value of any of
the perturbed quantities, which are shown in
Figure \ref{fig:perturbed_evolution}. As expected from the local linear mode analysis,
the amplitudes of the various components of the perturbation grow exponentially at
the same rate.

\begin{figure}
\centering
\includegraphics{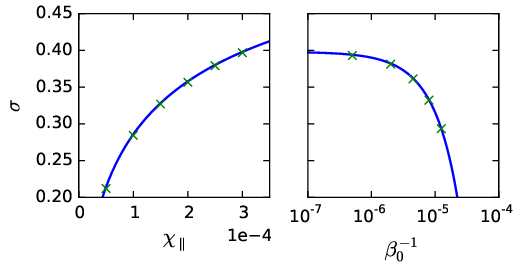}
\caption{Growth rates for the MTCI. \emph{Left}: The growth rate increases with
the value of $\chi_\para$. \emph{Right}: The growth rate decreases with
increasing initial magnetic field strength. The solid blue lines represent the
theoretical values evaluated at $z = L_z/2$. The green crosses are growth
rates obtained from the simulations.}
\label{fig:local_mtci}
\end{figure}

The growth rate of the MTCI depends on, among other things, the value of the
heat conductivity, $\chi_{\para}$, and the initial magnetic field strength,
$B_0$. In order to illustrate this dependence, and at the same time test our
modification to the code, we perform a parameter study. In the left panel of
Figure \ref{fig:local_mtci}, we show how the growth rate increases with the
value of the heat conductivity, $\chi_{\para}$. This is to be expected because
the MTCI is driven by heat transfer along magnetic field lines. In the right
panel of Figure \ref{fig:local_mtci}, we show how the growth rate decreases
with the value of $\beta_0^{-1}$. The explanation
for this behavior is that magnetic tension tends to stabilize the MTCI
\citep{Ber15a}. Magnetic tension has stabilizing effects in the limit
$\oma \gg \omdy$, where $\oma = k_\para v_{\rm A}$ and $\omdy = \sqrt{g/H_0}$.
In dimensionless units, this requirement can be written as $2k_\para^2 \gg
\beta$. From this estimate, the growth rates shown in the right panel of
Figure \ref{fig:local_mtci} should be negligible when $\beta_ {0}^
{-1} \gg 10^ {-4}$. The simulations and the solution to the dispersion
relation show that the growth rates are already inhibited by magnetic tension
at lower values of $\beta_{0}^{-1}$. These examples were generated by running
10 simulations at a modest resolution ($64\times 64$). At this resolution the
growth rates match to within a percent of the values expected from linear
theory.

\pagebreak

\subsection{The Heat and Particle-flux-driven Buoyancy Instability}
\label{sec:The Heat and Particle-flux-driven Buoyancy Instability}

When the magnetic field is parallel to gravity, the general dispersion relation
reduces to
\be
\sigma^{2}\approx g\frac{d\ln\left(T\mu\right)}{dz}\frac{k_{x}^{2}+
k_{y}^{2}}{k^{2}}\label{eq:HBCI_cond} \ ,
\en
in the limit of fast heat conduction and weak magnetic field.
The isothermal atmosphere where $\mu$ increases with height, that we
considered in the previous section, is therefore also unstable when
the magnetic field is vertical. In this case, the instability has been termed
the HPBI \citep{Pes13}.

\begin{figure}
\centering
\includegraphics{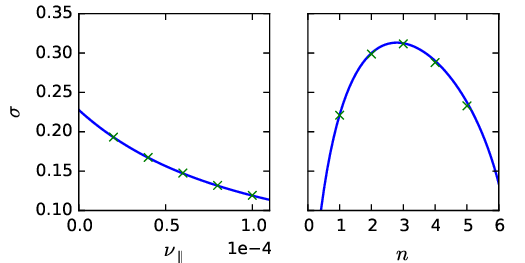}
\caption{Growth rates for the local HPBI. \emph{Left}: The growth rate
decreases with the value of $\nu_\para$. \emph{Right}: The growth rate as
a function of the mode number, $n = n_x = n_z$. The solid blue lines represent
the theoretical values evaluated at $z = L_z/2$. The green crosses are growth
rates obtained from the simulations.}
\label{fig:local_hbci}
\end{figure}

In this example, we include Braginskii viscosity which inhibits the growth rate by damping
perturbations perpendicular to the magnetic field. Braginskii viscosity can be
important for the HPBI \citep{Ber15a}. The mechanism is similar to the
mechanism described by \cite{Kun11} for the HBI. In order to excite a single
mode of the HPBI, we use a perturbation with wavenumbers $k_x = 2\pi n_x/L_x$
and $k_z = 2\pi n_z/L_z$, where $n=n_x=n_z$ is the mode number. We show the
growth rate
as a function of the Braginskii viscosity coefficient, $\nu_\para$ in the left
panel of Figure \ref{fig:local_hbci}. These simulations used a fixed value of
$\chi_\para = 10^{-4}$, $n = 1$ and a numerical resolution of $64 \times 64$.
As expected,  the growth rate indeed decreases with increasing value of
viscosity $\nu_{\para}$.

The second dependence we study for the HPBI is the one on the mode number,
$n$. High wavenumbers require higher numerical resolution in order to be
resolved and we use a resolution of $256\times256$ for these simulations. For
the sake of simplicity, Braginskii viscosity is not included in these
simulations. The result is shown in the right panel of Figure
\ref{fig:local_hbci}. The growth rate increases for increasing wave number
because small wavelength perturbations have a shorter time scale for heat
conduction.  When the wavelength is too short magnetic field tension renders
the modes stable. A naive estimate, using $2k_\para^2 \gg \beta$,
suggests that this should happen when $n \gg 16$, but the exact solution to
the dispersion relation shows that the instability is quenched already when $n
= 7$. Using such simulations we can directly see the cutoff in unstable wave
numbers resulting from magnetic field tension (as in this case) or viscosity
(not shown here).

\subsection{Modes Driven by Diffusion}
\label{sec:Modes Driven by Diffusion}

One of the interesting findings of \cite{Pes13} is that there are
instabilities that are driven by particle diffusion. This means that even
though the equilibrium is stable according to Equation (\ref{eq:HBCI_cond}),
the fact that $D\neq 0$ makes the equilibrium unstable. In order to study
these unstable modes, we assume, for simplicity,
an isothermal atmosphere with an initially vertical magnetic field.

\begin{figure}[t]
\centering
\includegraphics{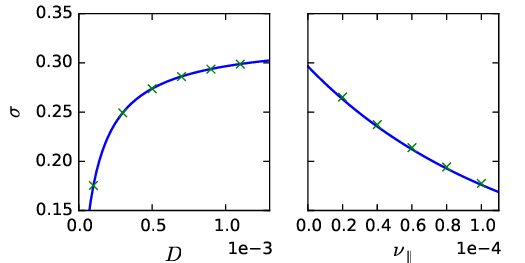}
\caption{Growth rate as a function of
$D$ (left) and $\nu_\para$ for fixed values of $\chi_\para = D =  10^{-3}$
(right). The solid blue line represents the theoretical values evaluated at
$z = L_z/2$ and the green crosses are growth rates obtained from the
simulations.}
\label{fig:difmode1}
\end{figure}

In this case, as explained in Section \ref{sec:Plane-parallel atmosphere with
gradients in temperature and composition}, an equilibrium configuration needs
to fulfill $\nabla\cdot\bb{Q}_{\rm c}=0$ and so we consider the atmosphere given in
Section \ref{sec:advanced_atmo} as initial condition. According to Equation (\ref{eq:HBCI_cond}), this configuration is
unstable to the HPBI for an isothermal atmosphere
when the Helium concentration increases vertically.
If instead the Helium concentration decreases with height, the
atmosphere is stable in the absence of anisotropic particle diffusion.
Choosing the slope in composition to be $s_{\rm c} = -0.01$, we do not observe any
instabilities in the simulation when $D = 0$.  The situation changes dramatically,
turning unstable when $D\neq  0$. The growth rates found in such simulations are compared with the
predictions from the linear theory in Figure \ref{fig:difmode1}. Since the
modes are driven by diffusion of Helium, we expect the growth rate to increase
with the value of $D$ (left panel). The modes have a damped growth rate when
Braginskii viscosity is included. We observe a decrease in the growth rate
with increasing $\nu_\para$, in agreement with the solution to the dispersion
relation (right panel).

\subsection{Gradients in temperature and composition}
\label{sec:Gradients in temperature and composition}

Having tested the case of isothermal atmospheres, we now consider a more
general situation where both $dT/dz \neq 0$ and $d\mu/dz \neq 0$.
In order to work with sensible values for these gradients, we  consider the
models in \cite{Pen09}, who analyzed the long-term evolution of the concentration
of Helium in a one-dimensional setting by solving a coupled set of Burgers' equations
for a multicomponent plasma in the absence of a magnetic field.
\cite{Ber15a} analyzed the stability of the \cite{Pen09} model by focusing on
local regions, characterized by fixed temperature and composition gradients,
and modeling these as a plane parallel atmosphere.

In this section, we present local simulations with gradients in temperature and
composition estimated at $r/r_{500} = 0.02$ and $r/r_{500} = 0.5$ with
$r_{500} = 1.63$ Mpc in the \cite{Pen09} model. These are the locations that were
analyzed in Section 6.6 and 6.4 in \cite{Ber15a}, indicated with a $C$ and an
$A$ in Figure 8 in that paper. These two locations correspond to the inner
region where the temperature and composition increase with radius and the
outer region where the temperature and composition decrease with radius. At
these radii, the values for the logarithmic gradients are $\dT = -0.4$ and
$\dmu = -0.13$ at  $r/r_{500} = 0.02$ and $\dT = 0.16$ and $\dmu = 0.05$ at
$r/r_{500} = 0.5$.

\begin{figure}
\centering
\includegraphics{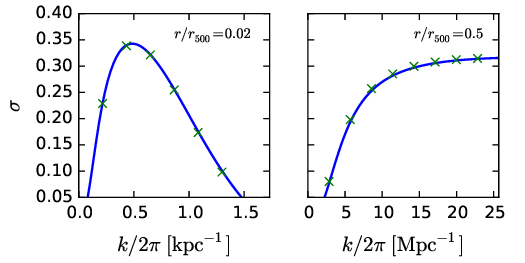}
\caption{\emph{Left:} Growth rates in the inner region as a function of
$k = k_x = k_z$. \emph{Right:} Growth rates in the outer region as a function
of $k_x$ for $k_z = 0$. The solid blue line represents the theoretical values
evaluated at $z = L_z/2$ and the green crosses are growth rates obtained from
the simulations.}
\label{fig:icm_linear}
\end{figure}

We use the equilibrium derived in Section \ref{sec:advanced_atmo} with
values taken from the model of \cite{Pen09},  $T_0 =
6.6$ keV ($T_0 = 9.5$ keV) and $c_0 = 0.56$ ($c_0 = 0.29$) for the inner
(outer) region. The computational domain is $L_x = L_z = H_0/10$ where $H_0 =
50$ kpc for the inner region and $L_x = 10 L_z = H_0$ where $H_0 = 0.35$ Mpc
for the outer region.
The gradients in composition and temperature are set such that the
dimensionless values of $\dT$ and $\dmu$ in the plane-parallel atmosphere agree with
the values in the model of \cite{Pen09}. They are given
by $s_{\rm c} = 2.9$
Mpc$^{-1}$ ($s_{\rm c}= -0.19$ Mpc$^ {-1}$) and $s_{\rm T}= 0.058$ keV kpc$^
{-1}$ ($s_{\rm T} = -4.3$ keV Mpc$^{-1}$) for the inner (outer) region.

The values for $\nu_\para$, $\chi_\para$, and $D$ are calculated from the
model of \cite{Pen09} as explained
in the Appendix of \cite{Ber15a}. The dimensionless values are so large that
high resolution numerical simulations become very computationally expensive.
As this is a test, we have arbitrarily reduced the values by a factor of
100 in the simulations. We use a value of $\beta = 2\cdot10^{6}$ for both
sets of simulations and adopt a resolution of $32\times 256$ (MTCI) and
$256 \times 256$ (HPBI). Some of the details of the simulations are listed in Table
\ref{tbl-1} with the names HPBI\_ICM and MTCI\_ICM. The growth rates
also depend on the wavenumbers, $k_x$ and $k_z$. For the HPBI (in the inner
region) we take $k = k_x = k_z$ and investigate growth rate as a function of
$k$. For the MTCI (in the outer region) we take $k_z = 0$ and investigate the
growth rate as a function of $k_x$. The results are shown in Figure
\ref{fig:icm_linear} with the growth rates of the HPBI in the left panel and
the growth rates of the MTCI in the right panel. An estimate shows
that the HPBI should be suppressed by magnetic tension for $k/2\pi \gg 3.4
\textrm{ kpc}^{-1}$ and the MTCI should be suppressed for $k/2\pi \gg 450
\textrm{ Mpc}^{-1}$. The growth rates are in units
of 50 and 280 Myr, respectively. Therefore, in physical units,  the maximum growth rates
in these simulations are  $\sigma_\mathrm{max} = 6.4 \ \mathrm{Gyr}^{-1}$ for the
HBPI and $\sigma_\mathrm{max} = 1.2 \ \mathrm{Gyr}^{-1}$ for the MTCI.

\begin{deluxetable}{lrrlllllrrrcrl}
\tabletypesize{\scriptsize}
\tablecaption{Overview of the simulations of the nonlinear regime using the
reflective boundary conditions.\label{tbl-2}}
\tablewidth{0pt}
\tablehead{
\colhead{Simulation} & $\theta$& \colhead{$\beta_0$} &
\colhead{$\chi_\para$} &\colhead{$\nu_\para$} & \colhead{$D$} & \colhead{Resolution}
& \colhead{Figure} }
\startdata
MTCI256      & $0^{\circ} $ & $2\cdot10^8$ & $5\cdot10^{-4}$ & 0 & 0 &
$256\times256$ &
\hyperref[fig:simple_nonlinear]{\ref{fig:simple_nonlinear}.a}, \ref
{fig:mtci_simple_nonlinear_energies}, \ref{fig:HPBI_magnetic_field_angle} \\
HPBI128      & $90^{\circ}$ & $2\cdot10^8$ & $5\cdot10^{-4}$ & 0 & 0 & $128\times128$ & \ref{fig:hpbi_simple_nonlinear_energies} \\
HPBI256      & $90^{\circ}$ & $2\cdot10^8$ & $5\cdot10^{-4}$ & 0 & 0 & $256\times256$ & \ref{fig:hpbi_simple_nonlinear_energies}\\
HPBI512      & $90^{\circ}$ & $2\cdot10^8$ & $5\cdot10^{-4}$ & 0 & 0 & $512\times512$ &
\hyperref[fig:simple_nonlinear]{\ref{fig:simple_nonlinear}.b}, \ref
{fig:hpbi_simple_nonlinear_energies}, \ref{fig:HPBI_magnetic_field_angle}
\enddata
\end{deluxetable}

\section{Simulations of the Nonlinear Regime}
\label{sec:Simulations of the Nonlinear Regime}
\begin{figure*}[t]
\includegraphics{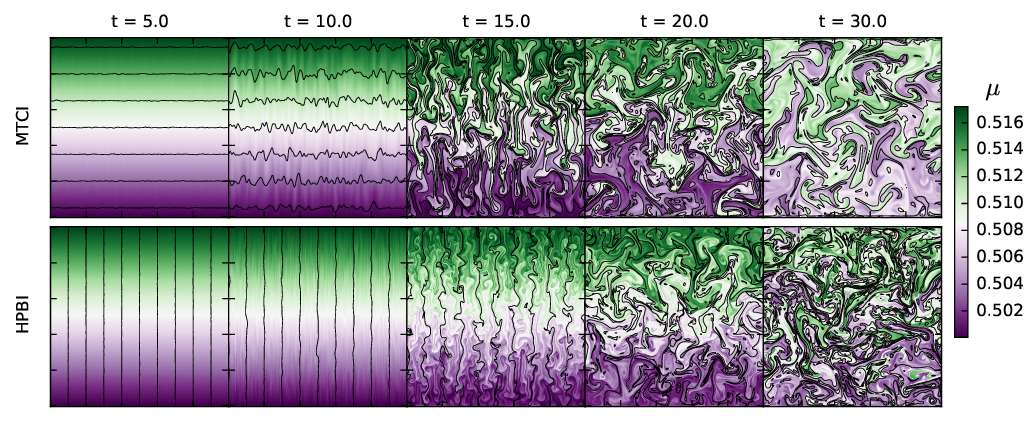}
\centering
\caption{Evolution of instabilities in an isothermal atmosphere with
$\dmu = -1/3$. The magnetic field lines are shown as solid black lines. The
composition of the plasma is shown with green representing a high
concentration and purple representing a low concentration. The MTCI
(upper panel) and the HPBI (lower panel) both give rise to mixing of
the Helium content. The size of the computational domain is $H_0/10
\times H_0/10$. The motions generated by the instabilities can be hinted
at by comparing neighboring snapshots but are best understood from the
animated version of this figure (see the online version).}
\label{fig:simple_nonlinear}
\end{figure*}

In order to study the nonlinear evolution of the MTCI and the HPBI we use the
reflective boundaries described in Appendix \ref{sec:Boundary
conditions_appendix}. We use the isothermal atmospheres presented in Section
\ref{sec:simple_atmo} and seed both velocity components with Gaussian noise
with a standard deviation of $10^{-4}$. The simulations are run without
Braginskii viscosity or anisotropic diffusion of Helium but anisotropic heat
conduction is accounted for with a value of $\chi_\para = 5\cdot10^{-4}$. We use a
value of $2\cdot10^{8}$ for the plasma-$\beta$. An overview of the simulations of
the nonlinear regime can be found in Table~\ref{tbl-2}.

\begin{figure}
\centering
\includegraphics{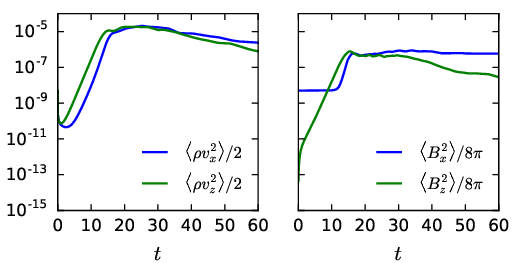}
\caption{Evolution of kinetic (left panel) and magnetic (right panel) energies
for the MTCI. After the initial phase of exponential growth, the instability
saturates with energies that are roughly in equipartition.
}
\label{fig:mtci_simple_nonlinear_energies}
\end{figure}

We start out by studying the evolution of the MTCI, i.e., we consider an atmosphere
threaded by a horizontal magnetic field. The subsequent
evolution of the magnetic field and the plasma composition is illustrated in
the upper panel of Figure \ref{fig:simple_nonlinear}. In this figure, it is
evident that the MTCI is able to mix the Helium content and to completely
rearrange the initially ordered magnetic field. The resulting growths in
kinetic and magnetic energy densities are shown, respectively, in the left
and right panels of Figure \ref{fig:mtci_simple_nonlinear_energies}.
The kinetic energies associated with the two velocity
components are roughly in equipartition throughout the simulation, i.e., $\langle \rho
v_x^{2} \rangle \approx \langle \rho v_z^{2} \rangle$ with $\langle \rho
v_z^{2} \rangle$ always larger but never exceeding $\langle \rho v_x^{2}
\rangle$ by more than an order of magnitude. The exponential phase of the
instability ends at $t \approx 30$. After this point in time, both the kinetic
and magnetic energies saturate with the former exceeding the latter
by two orders of magnitude.
In spite of the fact that  $\langle B_z^2\rangle$ vanishes initially, by the
end of the simulation the energies associated with the two magnetic field
components are roughly in  equipartition with $\langle B_z^2\rangle$ larger
than $\langle B_x^2\rangle$ by a factor of $\approx 2$, with $\langle B_x^2\rangle$
having grown by a factor of $\approx 8$ with respect to its initial value.

\begin{figure}
\centering
\includegraphics{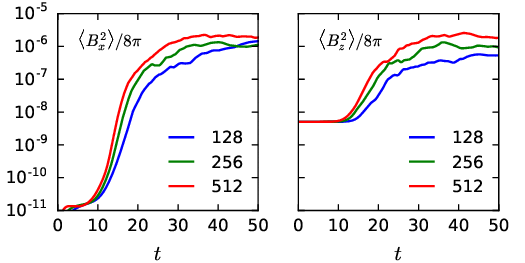}
\caption{Convergence of $\langle B_x^2 \rangle/8\pi$ and $\langle B_z^2
\rangle/8\pi$ as a function of resolution. The highest resolution is much more
expensive to run because of the prohibitive time step constraint due to heat
conduction, see Appendix \ref{sec:Implementation of Helium in Athena}.}
\label{fig:hpbi_simple_nonlinear_energies}
\end{figure}

We now consider the evolution of the HPBI. The setup is essentially the same
but the initial magnetic field is now vertical.\footnote{The boundary
conditions on the magnetic field are also slightly different, see Appendix
\ref{sec:Boundary conditions_appendix}.} The evolution of the HPBI is
illustrated in the lower panel of Figure \ref{fig:simple_nonlinear} with a
resolution of $512\times 512$. The initial vertical magnetic field is
rearranged by the HPBI, and, as for the MTCI, the Helium content is mixed by
the action of the instability. The HPBI leads to growth in the magnetic and
kinetic energy densities. In order to asses whether this growth is numerically
converged, we have also run simulations at resolutions of $128\times128$ and
$256\times256$. We show the evolution of $\langle B_x^2\rangle/8\pi$ and $\langle
B_z^2\rangle/8\pi$ for the three different numerical resolutions in Figure
\ref{fig:hpbi_simple_nonlinear_energies}. We observe that the instability
leads to exponential growth followed by saturation in both $\langle
B_x^2\rangle/8\pi$ and $\langle B_z^2\rangle/8\pi$. While the growth rate increases
with increasing resolution the values in the saturated state agree quite well.

It is also of interest to understand how the magnetic field changes from being initially
vertical to having a large horizontal component. The reason being the
consequences for heat transport along the vertical direction of the box. Such
studies have been done for both the MTI \citep{Par05,Par07} and the HBI
\citep{Par08}. These studies were motivated by a need to understand the
cooling flow problem of galaxy clusters \citep{Fab94}, and whether
magnetic fields could alleviate this problem. While the MTI could
potentially increase heat transport towards the core by making the magnetic
field be preferentially in the radial direction \citep{Par08_MTI}, the HBI has
been shown to lead to core insulation by driving the magnetic field to be
perpendicular to the radial direction \citep{Par08,Bog09,Par09}, which
would exacerbate the cooling flow problem.

\begin{figure}
\centering
\includegraphics{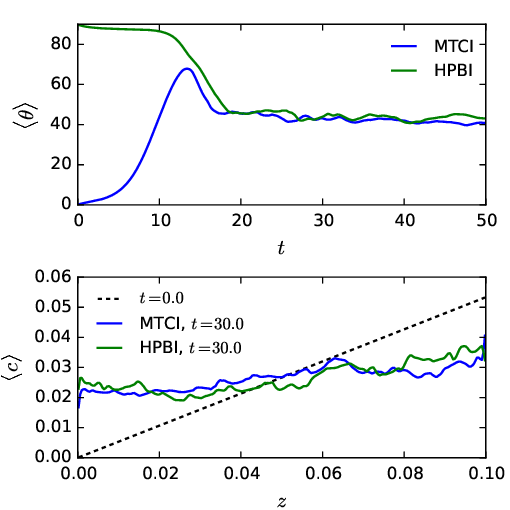}
\caption{\emph{Upper panel:} Evolution of the average inclination of the
magnetic field for the MTCI and the HPBI. Both instabilities seem to
drive the average inclination towards $45^{\circ}$. \emph{Lower panel:} The
average along $x$ of $c$ for the MTCI (blue) and the HPBI (green) at the end
of the exponential phase of the simulations ($t = 30$). The initial gradient
in $c$ (dashed black line) is diminished by the instabilities.}
\label{fig:HPBI_magnetic_field_angle}
\end{figure}

In Figure \ref{fig:HPBI_magnetic_field_angle}, we show the average
magnetic field
inclination as a function of time for the simulations of the MTCI and the
HPBI. The average inclination saturates to a value of approximately $\theta
\approx 45 ^{\circ}$ for both the simulations. This behavior is qualitatively
different from the behavior of the magnetic field inclination for the MTI and
the HBI. The difference can be explained in the following way. The MTI, which
is maximally unstable when the magnetic field is horizontal, has been found to
drive the saturated magnetic field to be roughly vertical \citep{Par07}. The HBI, which is
maximally unstable when the magnetic field is vertical, drives the magnetic
field to be roughly horizontal \citep{Par08}. These instabilities depend on gradients
in temperature that have opposite directions and so they cannot be present at
the same time.
On the other hand, both the MTCI and the HPBI require a mean
molecular weight that increases with height, and so they can both be present
at the same time. This feature of the MTCI and the HPBI was discussed in
\cite{Ber15a}, see especially Figure 4 in that paper. The interpretation of
the left panel of Figure \ref{fig:HPBI_magnetic_field_angle} is therefore that
the MTCI aims at driving the magnetic field angle towards $90^{\circ}$ while
the HPBI aims at driving the magnetic field angle towards $0^{\circ}$.
In the end, they reach a compromise at roughly $45^{\circ}$.

The Helium mass concentration, $c$, dramatically changes and the initial
gradient is diminished by the instability as time progresses. This is
illustrated in the lower panel of Figure \ref{fig:HPBI_magnetic_field_angle}
for both the MTCI and the HPBI. As explained in the introduction, gradients in
composition can introduce biases in key cluster parameters. We are therefore
interested in understanding whether such gradients, if initially present, will
be robust. The simulations presented here are heavily idealized, among many
reasons because the gas is assumed to be initially isothermal and the
simulations are local. Nevertheless, these simulations serve as a
proof-of-principle that gradients in composition can indeed be altered by
turbulent
mixing induced by plasma instabilities. Future work, using realistic gradients
for temperature and composition as well as transport coefficients should allow
us to understand whether such mixing can occur on timescales relevant for
galaxy clusters.

\pagebreak
\section{Summary and discussion}
\label{sec:Summary and discussion}

In this paper we have introduced a modified version of Athena
\citep{stone_athena:_2008} for performing kinetic MHD simulations of weakly
collisional plasmas with non-uniform composition. We have employed this
modified code to perform the first simulations of the MTCI, the HPBI and the
diffusion modes introduced in \citet{Pes13}. The set of simulations, aimed
at investigating the linear evolution of these instabilities, served as a test
for both the modification to Athena and the local linear mode analysis in
\cite{Pes13} and \cite{Ber15a}.

The simulations of weakly collisional, isothermal atmospheres with a gradient
in Helium presented in Section \ref{sec:Simulations of the Nonlinear Regime}
showed that the plasma instabilities, feeding off gradients in composition, can
induce turbulent mixing of the Helium content. This conclusion is valid for
compositions that increase in the direction anti-parallel to gravity,
regardless of whether the initial magnetic field is parallel or perpendicular
to the direction of gravity.
In the saturated state, the magnetic field components in the $x$ and $z$ directions
have roughly the same average energy but the energies are a factor of 10
higher for the HPBI than for the MTCI. The kinetic energy components are also
roughly in equipartition. In both cases, the instabilities saturate by
driving the average magnetic field inclination to roughly $45^{\circ}$.
This effect seems to open the possibility of alleviating the core insulation
observed in previous homogeneous simulations of the HBI. This is provided
that the global cluster dynamics were to allow for an increase in the mean
molecular weight with radius in the inner region, as envisioned by current
(one-dimensional, unmagnetized) Helium sedimentation models \citep{Pen09}.

The simulations of the nonlinear regime of the MTCI and the HBPI presented in
this paper considered an isothermal atmosphere as the equilibrium background.
It would be an improvement to use the model of \cite{Pen09} to determine the
gradients in both temperature and composition. This would provide insight into
the saturation of instabilities in potentially more realistic scenarios where the
dynamical evolution is determined by the simultaneous effects of both
gradients. Before proceeding with this endeavor there are, however, a
few issues that would be desirable to address, as we detail below.

It was found in \cite{Ber15a} that the HPBI, if present in the inner regions
of the ICM model of \cite{Pen09}, will have its fastest growth rates at
wavelengths that are longer than the scale height of the atmosphere. This
conclusion is similar to what was found for the HBI in \cite{Kun11}. Neither local
linear theory nor local simulations will therefore capture the physics of the
HPBI in the inner region of the ICM. This implies that both a quasi-global theory
and simulations are needed in order to study the influence of the possible gradient
in composition on the dynamics of the inner region of the ICM.

Local simulations of the MTI have been shown to underestimate the turbulence
\citep{2011MNRAS.413.1295M} and boundary effects can also modify the
conclusions from local simulations. The solution to this problem for the MTI
has been to sandwich the unstable region between stable layers, thereby
isolating it from the boundaries \citep{Par05,Par07,Kun12}. A similar approach
seems reasonable for the MTCI.

Other complications stem from the fact that pressure anisotropies,
shown to be important for the evolution of the MTI and HBI \citep
{Kun11,Kun12}, can give rise to microscale instabilities such as the
firehose and mirror instability \citep{schekochihin_fast_2006,Sch10}. These
small-scale instabilities are only excited once the pressure anisotropy grows
beyond $|p_\para -p_\perp|/P \gtrsim \beta^{-1}$. They do not appear in the
tests of the linear regime of the MTCI, HPBI and the diffusion modes
presented in Section \ref{sec:Simulations of the Linear Regime} because we
terminate the simulations before the stability criterion is violated. They are
not present in the simulations of the nonlinear regime in Section
\ref{sec:Simulations of the Nonlinear Regime} because we take the pressure to
be isotropic in these simulations (no Braginskii viscosity).
The problem with microscale instabilities is that they are not correctly
described by the framework of kinetic MHD \citep{schekochihin_plasma_2005}, an
issue that will need to be addressed for simulations of the nonlinear
evolution of the MTCI and the HPBI when Braginskii viscosity is included (see
\citealt{Kun12} for a discussion of these issues in the context of the MTI and
the HBI).

All this being said, our study suggests that, at least in the idealized settings
that we considered here, gradients in composition are able to drive turbulent
mixing of the composition in weakly collisional, magnetized plasmas.
This motivates future work on the generation and sustainment of both temperature
and composition gradients in galaxy clusters and their potential influence on the
global dynamics of the ICM. We envision that the modified version of Athena that we
developed will be a useful asset in this context.
In order to model more realistically the physics of the ICM, future
improvements could include extending the simulations to three
dimensions and adding optically thin cooling in order to study the cooling
flow problem. Furthermore, the equations of kinetic MHD, as embodied in
Equation (\ref{eq:rho})-(\ref{eq:c}), cannot account for the slow sedimentation
of Helium that is the core feature in the model of \cite{Pen09}.
An extension of the framework of kinetic MHD to include this effect would
allow us to self-consistently include sedimentation in the simulations
\citep{1990ApJ...360..267B,Ber15a} and study the effects of the instabilities
described in this paper in a dynamic, slowly varying background.

\acknowledgments

We acknowledge useful discussions with Daisuke Nagai, Matthew Kunz, Prateek
Sharma, Ellen Zweibel, and Ian Parrish during the \emph{3\textsuperscript{rd}
ICM Theory and Computation Workshop} held at the Niels Bohr Institute in 2014.
We thank Sagar Chakraborty, Colin McNally, Gareth C. Murphy, and
Henrik Latter for valuable discussions and comments.
We are grateful to Oliver Gressel for suggesting using the quasi-periodic
boundary conditions for testing the linear theory and to Tobias Heinemann for
aid in rendering magnetic field lines. We also thank the
anonymous referee for a number of useful suggestions that helped improve
the manuscript. The research leading to
these results has
received funding from the European Research Council under the European Union's
Seventh Framework Programme (FP/2007-2013) under ERC grant agreement 306614.
T. B. also acknowledges support provided by a L{\o}rup Scholar Stipend and M.
E. P. also acknowledges support from the Young Investigator Programme of the
Villum Foundation.

\appendix

In this appendix, we describe the numerical methods used in this paper. We use
the publicly available MHD code Athena which solves the conservative form of
the MHD equations. The algorithms used are described in
\cite{gardiner_unsplit_2005, stone_simple_2009} and the implementation of
Athena along with tests is described in detail in \cite{stone_athena:_2008}.
Athena is a finite volume code, which uses the Godunov method. We use the
directionally unsplit corner transport upwind method along with constrained
transport (CTU + CT) which is the recommended setting. We furthermore use the
anisotropic heat conduction module that was implemented in Athena by
\cite{Par05} using operator splitting.

Appendix \ref{sec:Implementation of Helium in Athena} explains the
implementation of a spatially varying mean molecular weight, $\mu$, the
anisotropic diffusion of Helium and tests cases, in  Athena.  In Appendix
\ref{sec:Boundary conditions_appendix} we discuss in detail the boundary
conditions used in the simulations.

\section{Implementation of Anisotropic Diffusion of Composition in Athena}
\label{sec:Implementation of Helium in Athena}

Let us consider the equation describing the evolution of the Helium
mass concentration, $c = \rho_{\rm He}/\rho$, given by\footnote{We do not
consider the effects of thermo-diffusion and baro-diffusion which makes our
current model unable to describe the slow sedimentation of Helium \citep
{1990ApJ...360..267B} that
can give rise to a composition gradient.}
\be
\D{c}{t} + \paren{\bb{v} \bcdot \del}c =
-\bb{\nabla}\bcdot\bb{Q}_{\rm c} \label{eq:contration_eqaution} \,.
\en
Athena has an option for adding passive scalars which we use for adding the
Helium mass concentration. This option turns on an extra equation
\be
\D{\paren{\rho c_{\rm n}}}{t} + \del \bcdot \paren{\rho c_{\rm n} \bb{v}} = 0 \ ,
\en
where $\rho$ is the total density and $c_{\rm n}$ is the mass concentration of the
$\textrm{n}^{\rm th}$ scalar. We only add a single scalar, namely the Helium mass
concentration, $c$. This built-in function takes care of the Lagrangian part
of Equation (\ref{eq:contration_eqaution}). The diffusion term is then solved
using a finite difference scheme and operator splitting.

Anisotropic diffusion of Helium is described by the RHS of Equation
(\ref{eq:contration_eqaution}), which, when $\bb{v} = 0$, reduces to
\be
\D{c}{t} = -\bb{\nabla}\bcdot\bb{Q}_{\rm c}  =
           D\del \bcdot \left(\b\b \bcdot \del c \right)\,.
\en
The composition flux for anisotropic Helium diffusion has
the same form as the heat flux for anisotropic heat conduction, as seen by
comparing Equations (\ref{eq:helium-flux}) and (\ref{eq:heat-flux}),
respectively. We can therefore use the same method to calculate the two
physically different anisotropic fluxes. The original implementation of
anisotropic heat conduction was done by \cite{Par05} using an asymmetric
finite difference scheme \citep{Sha07,vanEs2014526}.

Non-ideal effects are computationally expensive because they are generally
described by parabolic operators which cannot be added to the hyperbolic
fluxes used in the Godunov scheme. The parabolic operators can be shown to
have a very prohibitive time step constraint \citep{Dur} for heat conduction
and concentration diffusion as given by, respectively,
\be
\Delta t_{\chi_\para}<\frac{b}{\kappa_\para}\frac{\left(\Delta x\right)^{2}}
{\gamma-1} \,,
\label{eq:heat_time_step}
\en
\be
\qquad
\Delta t_D < b\frac{\left(\Delta x\right)^{2}}{D} \,.
\label{eq:diff_time_step}
\en
Here, $\kappa_\para = \chi_\para T/P$ is the heat diffusivity,
the parameter $b$ is $b=1/2,\,1/4,\,1/6$ in one, two and
three dimensions, respectively, and $\Delta x$ is the grid size.

The Courant number, $C$, is defined to be the ratio of the applied time step
to the allowed time step. We use $C = 0.4$ in all our simulations. Because
$\Delta t_{\rm MHD} \propto \Delta x$, the very prohibitive constraints on the
time step for the parabolic operators will generally lead to $\Delta t_{\rm
MHD} \gg \Delta t_\chi  \sim \Delta t_D$. In order to partially circumvent
this problem we use subcycling, taking up to ten diffusion steps for each MHD
step, as suggested in \citep{Kun12}.

\cite{Sha07} found that the finite difference approximation can lead to
unphysical behavior with diffusion in the wrong direction. In the context of
heat diffusion this problem can lead to negative temperatures and therefore
an imaginary sound speed. The same problem arises when considering Helium
diffusion and we use Van
Leer limiters on the derivatives to circumvent it \citep{Sha07}.

The publicly available version of Athena  works with constant viscosity,
$\nu_\para$, and heat diffusivity, $\kappa_\para = \chi_\para T/P$. However,
these coefficients, as well as the diffusion coefficient $D$, do in general
depend on temperature, density and composition, see for instance, the
discussion in the Appendix of \cite{Ber15a}.
Accounting for this dependence is not crucial in local
simulations but it becomes essential in global simulations. We have modified
Athena to use spatially varying coefficients by using a harmonic average of
the coefficients \citep{Sha07}. This makes the time step computed from
Equations (\ref{eq:heat_time_step}) and (\ref{eq:diff_time_step}) spatially
dependent.  We therefore calculate the time step at each cell and use the
minimum value. This implementation will be useful in future global studies.

\subsection{Tests of the implementation of anisotropic diffusion}

In order to verify the implementation of anisotropic diffusion of Helium,
we performed three different test problems with a known analytical
solution. These tests were carried out with the MHD solver turned off.

\subsubsection{One-dimensional diffusion}

\begin{figure}[t]
\centering
\includegraphics{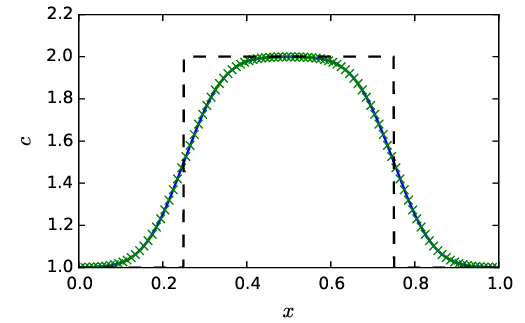}
\caption{Diffusion of a step function. The initial condition is shown with a
dashed line. The green crosses correspond to data from the simulation and the
solid blue line is the analytical solution at $t= 0.0028$.}
\label{fig:step-function}
\end{figure}

We consider the diffusion of a step function as in \cite{Ras08} using a one-dimensional
grid with 100 cells on the domain $x = [0,1]$ with $D = 1$ and run
the simulation up to $t= 0.0028$. The analytical solution to the diffusion of
a step function is \be c(x,t)=c_0+\frac{\Delta c}{2} \textrm{erf}\left(\pm
\frac{x-x_0}{\sqrt{4D t}}\right) \ , \en where $c_0 = 3/2$ and $\Delta c = 1$.
The $"+"$ sign is used with $x_0 = 0.25$ for $x<0.5$ and the $"-"$ sign is
used with
$x_0 = 0.75$ for $x>0.5$. The numerical result matches the analytical solution,
as seen in Figure \ref{fig:step-function}, implying that the method works
well in one dimension.

\subsubsection{Diffusion of a two-dimensional Gaussian}

A more challenging test can be posed by considering the magnetic field
to be inclined at an angle, $\theta$, with respect to the grid. We consider an initially
isotropic, two dimensional Gaussian distribution of Helium diffusing out along
an inclined magnetic field. The analytical solution is\footnote{This result
can be derived by solving the one-dimensional diffusion equation for a
Gaussian initial distribution followed by a rotation of the coordinate system.
The one-dimensional problem is solved by using a Fourier transform in space
and a Laplace transform in time.}
\be
c\left(x,y,t\right)=\frac{1}{2\pi a\left(t\right)a_{0}}\exp\left\{
-\frac{\left(x\cos\theta+y\sin\theta\right)^{2}}{2a\left(t\right)^{2}}\right\}
\nonumber \\
\times \exp\left\{-\frac{\left(y\cos\theta-x\sin\theta\right)^{2}}{2a_{0}^{2}}
\right\},\nonumber \\
\en
where $a(t)^2 = a_0^2 + 2 D t$ and $a_0$ it the initial standard deviation of
the Gaussian.

\begin{figure}[t]
\centering
\includegraphics{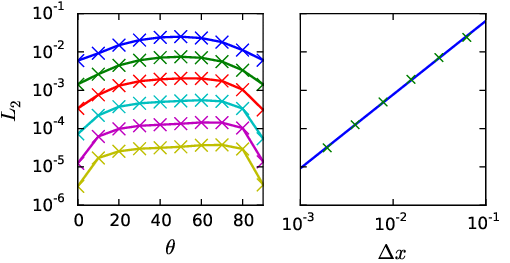}
\caption{\emph{Left:} $L_2$ error as a function of magnetic field inclination.
Resolutions of $N \times N $ with $N = 32$, $64$,
$128$, $256$, $512$ and $1024$ was used with monotonically decreasing $L_2$ at
all angles. As expected, the asymmetric finite difference scheme gives the
best result when the magnetic field is aligned with the grid. \emph{Right:}
Convergence to the exact solution with decreasing $\Delta x$ for a magnetic
field inclined at $40^\circ$ from the $x$-axis.}
\label{fig:Gaus2D-errors}
\end{figure}

The computational domain is a $[-1, 1] \times [-1, 1]$ Cartesian box. We use
$a_0 = 1/8$ and $D = 0.001$. The errors at $t= 4$ are compared to the
analytical solution in Figure \ref{fig:Gaus2D-errors}. In the left panel
the $L_2$ errors are shown as a function of the magnetic field
inclination and resolution. These errors are smallest when $\theta = 0$ or
$\theta = \pi/2$, corresponding to the grid and the magnetic field being
aligned. In the right panel we show that the solution for $\theta =
40^{\circ}$ converges as $ L_2 \propto (\Delta x)^m \ , $ where $\Delta x$ is
the (uniform) grid spacing and $m = 1.9$ is the order of convergence.

\subsubsection{Diffusion of a High-concentration Patch in a Circular Magnetic Field}

The final and most challenging test that we carry out for anisotropic transport
was introduced in \cite{Par05}. We consider a Cartesian box of size
$[-1, 1] \times [-1, 1]$ with a patch with higher concentration $c$,
specifically\footnote{This test was constructed for the anisotropic heat
conduction. We are using the same initial values ($10$ and $12$) as in the
literature, making it easier to compare the results. These values are of
course not meaningful values for $c$ but it still serves as a test of the
implementation of anisotropic diffusion. The same considerations apply to the
step function test.}
\be
c =\left\{ \begin{array}{cc}
12 & \quad \mbox{if} \quad 0.5<r<0.7 \quad \mbox{and} \quad
-\pi/12<\theta<\pi/12 \,, \\
10 & \mbox{otherwise} \,.
\end{array}
\right. \nonumber \\
\en
The density is uniform
with $\rho = 1$ and the magnetic field is circular. In order to ensure $\del
\bcdot \bb{B} = 0$, the magnetic field was initialized with a vector potential
satisfying $\del \btimes \bb{A} = \bb{B}$.

\begin{figure}
\centering
\includegraphics{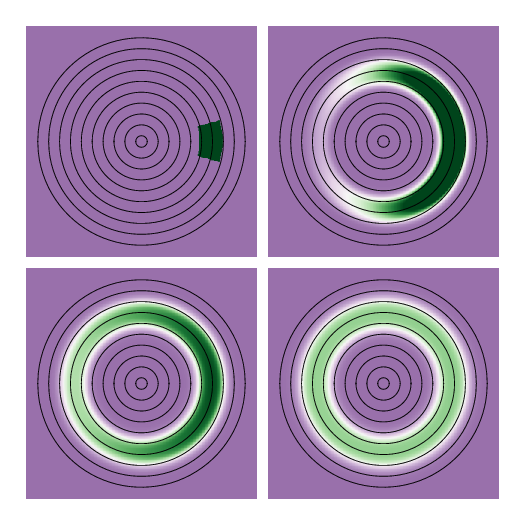}
\caption{The patch with a high concentration of $c$ diffuses out along the
circular magnetic field. Snap shots at $t = 0$, $25$, $75$ and $200$.
The color scale is from 10 to 10.2  and the perpendicular diffusion is small.
The resolution in this numerical experiment is $400\times 400$.}
\label{fig:HotPatch}
\end{figure}

We considered the value $D = 0.01$ and run the simulation until $t = 200$. The
over concentration diffuses out along the magnetic field lines, as observed in
Figure \ref{fig:HotPatch}. We have run this test problem with the same
resolutions as \cite{Sha07} and obtain the exact same values quoted there
for the error norms associated with the resolutions $200\times 200$ and $400
\times 400$. For instance, for a
resolution of  $200\times200$, we obtain $L_1 = 0.0264$, $L_2 = 0.0407$,
$L_\infty = 0.0928$, $c_{\rm min} = 10$ and $c_{\rm max} = 10.1016$ at
$t=200$ as stated in \cite{Sha07}.

It is evident from Figure \ref{fig:HotPatch} that,
even though only anisotropic diffusion is explicitly turned on, there is still a small
amount of numerical, perpendicular diffusion. This is undesired in simulations
of instabilities because isotropic diffusion will lower the growth rates or
even quench the instabilities. This was investigated by \cite{Par05} for the
MTI who found that it is, however, insensitive to perpendicular diffusion
provided that $\chi_\perp/\chi_\para < 10^{-3}$. The fact that we find the correct
analytical growth rates for all simulations discussed in Section
\ref{sec:Simulations of the Linear Regime} shows that the numerical
perpendicular diffusion is not a problem for our present purposes.

\section{Boundary conditions}
\label{sec:Boundary conditions_appendix}

In the horizontal direction we use the periodic boundary conditions that
Athena provides as a standard option. In the vertical direction, boundary
conditions that maintain hydrostatic equilibrium are required. In this section
we describe the conventional, reflective boundary conditions as well as the
quasi-periodic boundary conditions alluded to in Section~\ref{sec:Boundary conditions}.

\subsection{Reflective boundary conditions}

Our implementation of reflective boundary conditions follows the description
in \cite{Zin02}. Hydrostatic equilibrium requires that Equation
(\ref{eq:hydrostatic_equilibrium}) is satisfied. This requirement can be approximated by
\be
P_{i}-P_{i+1}=\frac{a g\, \Delta z}{12}\left(5\rho_{i+1}+8\rho_{i}
-2\rho_{i-1}\right) \ , \label{eq:hydrostatic_finite}
\en
where $a = 1$ ($a = -1$ ) at the top (bottom) of the domain. In this notation,
$i$ refers to cell $i$ and $i+1$ refers to one cell further up (down) when $a=
1$ ($a = -1$). This equation is then solved for $\rho_{i+1} $ using that
\be
P_{i+1} = \f{\rho_{i+1} T_{i+1}}{\mu_{i+1}} \ , \label{eq:idealgas_finite}
\en
along with an assumption on $\mu_{i+1}$ and $T_{i+1}$. One can either
assume
$\mu_i = \mu_{i+1}$ and $T_i = T_{i+1}$ or one can prescribe
the values at the boundaries to be equal to their initial values, i.e., $\mu_{i+1} = \mu_{0}$
and $T_{i+1} = T_{0}$.
In the case where the mean molecular weight $\mu$ is not included,
\cite{Par05} refer to these boundary conditions as adiabatic and conductive,
respectively. A combination of these two boundary conditions is also possible (i.e., fixing
$\mu$ and varying $T$ or vice versa). We have implemented all four
combinations but will only discuss the conducting boundary conditions
($\mu_{i+1} = \mu_{0}$ and $T_{i+1} = T_{0}$) in the following, since these
are the boundary conditions used in Section \ref{sec:Simulations of the
Nonlinear Regime}.

Solving Equations (\ref{eq:hydrostatic_finite}) and (\ref{eq:idealgas_finite})
for $\rho_{i+1}$ and $P_{i+1}$ we find
\be
\rho_{i+1} = \frac{P_{i}+ \alpha \paren{8\rho_i - \rho_{i-1} }}{T_{0}/\mu_{0}
- 5\alpha} \ ,
\en
and $P_{i+1} = \rho_{i+1} T_0 /\mu_0$, where
\be
\alpha = \f{a \Delta z \,g}{12}.
\en
These relations are used to calculate the density and pressure of the four
ghost zones at the top and bottom of the computational domain. At the same
time, velocity is reflected symmetrically around $z = 0$ and $z = L_z$. The
magnetic field components are also mirrored. In the case of initially vertical
magnetic field, we let the $B_x$ component change sign, whereas in the case
of initially horizontal magnetic field, we let the $B_z$ component change sign.
This forces the field to remain vertical (horizontal) at the boundary in the
case of initially vertical (horizontal) field.

Athena uses the Godunov scheme which is known not to be optimal at maintaining
hydrostatic equilibrium \citep{Zin02}. The reason is that the pressure term in
the momentum equation is not solved simultaneously with the gravity term.
There are ways to modify a Godunov scheme such that this problem is
circumvented, see for instance \cite{Zin02,Kap14}. We use a high numerical
resolution and a low Courant number ($C = 0.4$) in order to maintain
hydrostatic equilibrium as well as possible. The minimum amplitude we can use
for perturbations in $\bb{v}$ is, however, limited by the numerical noise
caused by the inability of the code to perfectly maintain hydrostatic equilibrium, in
agreement with the findings of \cite{Par12}.

\subsection{Quasi-periodic boundary conditions}

A key assumption in standard local linear mode analysis, such as
presented in \cite{Pes13,Ber15a}, is that the perturbations have the  spatial
dependence $\exp (i \bb{k \cdot x})$. This assumption is not fulfilled for the
reflective boundary conditions and it is thus impossible to cleanly excite a
single eigenmode. The problem being that the boundary conditions
excite other modes in an uncontrolled way. We originally realized this problem
when we studied the HBI but it persists in the case of the HPBI and its
diffusive variant. The problem is not present for the MTI and the MTCI in the
case $k_z = 0$ because the boundaries are periodic in $x$.

We have developed special boundary conditions that are consistent with the
assumptions used in the local mode analysis. One of the key assumptions here
is  that the perturbed quantities $ \delta v_x$, $\delta v_z$, $ \delta B_x$,
$\delta B_z$, $\delta \rho / \rho$, $\delta \mu / \mu$ and $\delta T / T$
are periodic.
In the following, the values outside the
computational domain (the ghost zones) are denoted by a subscript $g$ and the
values on the inside are denoted by a subscript $i$. The subscript
$"\textrm{eq}"$
refers to the value of the equilibrium background (as given in Section
\ref{sec:simple_atmo} or Section \ref{sec:advanced_atmo}).
The mapping from interior to ghost zones ($i \rightarrow g$) is the same as for periodic boundary conditions.
Instead of directly mapping the interior values to the ghost zones, we let the
ghost zones depend on the change in the interior values with respect to the
equilibrium background. The quasi-periodic boundary conditions are then defined as
\be
\rho_{g}&=& \rho_{g,\textrm{eq}}\left(1+\frac{\rho_i-\rho_{i,\textrm{eq}}}
{\rho_{i,\textrm{eq}}}
\right)
\ , \\
T_{g}    &=&T_{g,\textrm{eq}}\left(1+\frac{T_i-T_{i,\textrm{eq}}}{T_{i,\textrm{eq}}} \right) \ , \\
\mu_{g}&=&\mu_{g,\textrm{eq}}\left(1+\frac{\mu_i-\mu_{i,\textrm{eq}}}{\mu_{i,\textrm{eq}}}  \right) \ ,
\en
with the pressure given by $P_g = \rho_g T_g/\mu_g$. The equilibrium magnetic
field and velocity do not have a gradient and so their boundary conditions are
simply periodic, i.e. $\bb{v}_{g} = \bb{v}_{i}$ and $\bb{B}_{g} = \bb{B}_{i}$.
These are the boundary conditions we used in Section \ref{sec:Simulations of the
Linear Regime}.

\bibliography{Bibliography}

\begin{thebibliography}{}
\expandafter\ifx\csname natexlab\endcsname\relax\def\natexlab#1{#1}\fi

\bibitem[{{Bahcall} \& {Loeb}(1990)}]{1990ApJ...360..267B}
{Bahcall}, J.~N., \& {Loeb}, A. 1990, \apj, 360, 267

\bibitem[{{Balbus}(2000)}]{Bal00}
{Balbus}, S.~A. 2000, \apj, 534, 420

\bibitem[{{Balbus}(2001)}]{Bal01}
---. 2001, \apj, 562, 909

\bibitem[{{Berlok} \& {Pessah}(2015)}]{Ber15a}
{Berlok}, T., \& {Pessah}, M.~E. 2015, \apj, 813, 22

\bibitem[{{Bogdanovi{\'c}} {et~al.}(2009){Bogdanovi{\'c}}, {Reynolds},
  {Balbus}, \& {Parrish}}]{Bog09}
{Bogdanovi{\'c}}, T., {Reynolds}, C.~S., {Balbus}, S.~A., \& {Parrish}, I.~J.
  2009, \apj, 704, 211

\bibitem[{Braginskii(1965)}]{Bra}
Braginskii, S. 1965, Review of Plasma Physics

\bibitem[{{Chuzhoy} \& {Loeb}(2004)}]{Chu04}
{Chuzhoy}, L., \& {Loeb}, A. 2004, \mnras, 349, L13

\bibitem[{{Chuzhoy} \& {Nusser}(2003)}]{Chu03}
{Chuzhoy}, L., \& {Nusser}, A. 2003, \mnras, 342, L5

\bibitem[{Durran(2010)}]{Dur}
Durran, D.~R. 2010, Numerical Methods for Fluid Dynamics (Springer)

\bibitem[{{Fabian}(1994)}]{Fab94}
{Fabian}, A.~C. 1994, \araa, 32, 277

\bibitem[{{Fabian} \& {Pringle}(1977)}]{Fab77}
{Fabian}, A.~C., \& {Pringle}, J.~E. 1977, \mnras, 181, 5P

\bibitem[{Gardiner \& Stone(2005)}]{gardiner_unsplit_2005}
Gardiner, T.~A., \& Stone, J.~M. 2005, Journal of Computational Physics, 205,
  509

\bibitem[{{Gilfanov} \& {Syunyaev}(1984)}]{Gil84}
{Gilfanov}, M.~R., \& {Syunyaev}, R.~A. 1984, Soviet Astronomy Letters, 10, 137

\bibitem[{Kunz(2011)}]{Kun11}
Kunz, M.~W. 2011, Monthly Notices of the Royal Astronomical Society, 417, 602

\bibitem[{Kunz {et~al.}(2012)Kunz, Bogdanovi{\'c}, Reynolds, \& Stone}]{Kun12}
Kunz, M.~W., Bogdanovi{\'c}, T., Reynolds, C.~S., \& Stone, J.~M. 2012, The
  Astrophysical Journal, 754, 122

\bibitem[{{Latter} \& {Kunz}(2012)}]{Lat12}
{Latter}, H.~N., \& {Kunz}, M.~W. 2012, \mnras, 423, 1964

\bibitem[{{Markevitch}(2007)}]{Mar07}
{Markevitch}, M. 2007, ArXiv e-prints, arXiv:0705.3289

\bibitem[{{McCourt} {et~al.}(2011){McCourt}, {Parrish}, {Sharma}, \&
  {Quataert}}]{2011MNRAS.413.1295M}
{McCourt}, M., {Parrish}, I.~J., {Sharma}, P., \& {Quataert}, E. 2011, \mnras,
  413, 1295

\bibitem[{{McCourt} {et~al.}(2012){McCourt}, {Sharma}, {Quataert}, \&
  {Parrish}}]{2012MNRAS.419.3319M}
{McCourt}, M., {Sharma}, P., {Quataert}, E., \& {Parrish}, I.~J. 2012, \mnras,
  419, 3319

\bibitem[{{Parrish} {et~al.}(2012{\natexlab{a}}){Parrish}, {McCourt},
  {Quataert}, \& {Sharma}}]{Par12}
{Parrish}, I.~J., {McCourt}, M., {Quataert}, E., \& {Sharma}, P.
  2012{\natexlab{a}}, \mnras, 422, 704

\bibitem[{{Parrish} {et~al.}(2012{\natexlab{b}}){Parrish}, {McCourt},
  {Quataert}, \& {Sharma}}]{2012MNRAS.419L..29P}
---. 2012{\natexlab{b}}, \mnras, 419, L29

\bibitem[{{Parrish} \& {Quataert}(2008)}]{Par08}
{Parrish}, I.~J., \& {Quataert}, E. 2008, ApJL, 677, L9

\bibitem[{{Parrish} {et~al.}(2009){Parrish}, {Quataert}, \& {Sharma}}]{Par09}
{Parrish}, I.~J., {Quataert}, E., \& {Sharma}, P. 2009, \apj, 703, 96

\bibitem[{{Parrish} {et~al.}(2010){Parrish}, {Quataert}, \&
  {Sharma}}]{2010ApJ...712L.194P}
---. 2010, ApjL, 712, L194

\bibitem[{Parrish \& Stone(2005)}]{Par05}
Parrish, I.~J., \& Stone, J.~M. 2005, The Astrophysical Journal, 633, 334

\bibitem[{Parrish \& Stone(2007)}]{Par07}
---. 2007, The Astrophysical Journal, 664, 135

\bibitem[{{Parrish} {et~al.}(2008){Parrish}, {Stone}, \&
  {Lemaster}}]{Par08_MTI}
{Parrish}, I.~J., {Stone}, J.~M., \& {Lemaster}, N. 2008, \apj, 688, 905

\bibitem[{Peng \& Nagai(2009)}]{Pen09}
Peng, F., \& Nagai, D. 2009, The Astrophysical Journal, 693, 839

\bibitem[{{Pessah} \& {Chakraborty}(2013)}]{Pes13}
{Pessah}, M.~E., \& {Chakraborty}, S. 2013, \apj, 764, 13

\bibitem[{Quataert(2008)}]{Qua08}
Quataert, E. 2008, The Astrophysical Journal, 673, 758

\bibitem[{{R. K{\"a}ppeli and S. Mishra}(2014)}]{Kap14}
{R. K{\"a}ppeli and S. Mishra}. 2014, Journal of Computational Physics, 259,
  199

\bibitem[{Rasera \& Chandran(2008)}]{Ras08}
Rasera, Y., \& Chandran, B. 2008, The Astrophysical Journal, 685, 105

\bibitem[{{Ruszkowski} \& {Oh}(2010)}]{2010ApJ...713.1332R}
{Ruszkowski}, M., \& {Oh}, S.~P. 2010, \apj, 713, 1332

\bibitem[{Schekochihin \& Cowley(2006)}]{schekochihin_fast_2006}
Schekochihin, A.~A., \& Cowley, S.~C. 2006, Astronomische Nachrichten, 327,
  599, {arXiv}: astro-ph/0508535

\bibitem[{Schekochihin {et~al.}(2005)Schekochihin, Cowley, Kulsrud, Hammett, \&
  Sharma}]{schekochihin_plasma_2005}
Schekochihin, A.~A., Cowley, S.~C., Kulsrud, R.~M., Hammett, G.~W., \& Sharma,
  P. 2005, The Astrophysical Journal, 629, 139

\bibitem[{{Schekochihin} {et~al.}(2010){Schekochihin}, {Cowley}, {Rincon}, \&
  {Rosin}}]{Sch10}
{Schekochihin}, A.~A., {Cowley}, S.~C., {Rincon}, F., \& {Rosin}, M.~S. 2010,
  \mnras, 405, 291

\bibitem[{Sharma \& Hammett(2007)}]{Sha07}
Sharma, P., \& Hammett, G.~W. 2007, Journal of Computational Physics, 227, 123,
  {arXiv}:0707.2616 [astro-ph, physics:physics]

\bibitem[{{Shtykovskiy} \& {Gilfanov}(2010)}]{Sht10}
{Shtykovskiy}, P., \& {Gilfanov}, M. 2010, \mnras, 401, 1360

\bibitem[{{Spitzer}(1962)}]{1962pfig.book.....S}
{Spitzer}, L. 1962, {Physics of Fully Ionized Gases}

\bibitem[{Stone \& Gardiner(2009)}]{stone_simple_2009}
Stone, J.~M., \& Gardiner, T. 2009, New Astronomy, 14, 139

\bibitem[{Stone {et~al.}(2008)Stone, Gardiner, Teuben, Hawley, \&
  Simon}]{stone_athena:_2008}
Stone, J.~M., Gardiner, T.~A., Teuben, P., Hawley, J.~F., \& Simon, J.~B. 2008,
  The Astrophysical Journal Supplement Series, 178, 137

\bibitem[{van Es {et~al.}(2014)van Es, Koren, \& de~Blank}]{vanEs2014526}
van Es, B., Koren, B., \& de~Blank, H.~J. 2014, Journal of Computational
  Physics, 272, 526

\bibitem[{Zingale {et~al.}(2002)Zingale, Dursi, ZuHone, Calder, Fryxell, Plewa,
  Truran, Caceres, Olson, Ricker, Riley, Rosner, Siegel, Timmes, \&
  Vladimirova}]{Zin02}
Zingale, M., Dursi, L.~J., ZuHone, J., {et~al.} 2002, The Astrophysical Journal
  Supplement Series, 143, 539

\end{thebibliography}

\end{document}